\documentclass{sigchi}

\usepackage{balance}  
\usepackage{graphics} 
\usepackage{times}    
\usepackage{url}      
\usepackage{ifpdf}

\makeatletter
\def\url@leostyle{%
  \@ifundefined{selectfont}{\def\UrlFont{\sf}}{\def\UrlFont{\small\bf\ttfamily}}}
\makeatother
\def\pprw{8.5in}
\def\pprh{11in}

\ifpdf
\fi

\ifpdf
\usepackage[pdftex]{hyperref}
\hypersetup{
pdftitle={SIGCHI Conference Proceedings Format},
pdfauthor={LaTeX},
pdfkeywords={SIGCHI, proceedings, archival format},
pdfstartview={FitH},
bookmarksnumbered,
colorlinks,
citecolor=black,
filecolor=black,
linkcolor=black,
urlcolor=black,
breaklinks=true,
}
\else
\usepackage{hyperref}
\hypersetup{
bookmarksnumbered,
colorlinks,
citecolor=black,
filecolor=black,
linkcolor=black,
urlcolor=black,
breaklinks=true,
}
\fi

\begin{document}

\title{Design Activism for Minimum Wage Crowd Work}

\numberofauthors{3}
\author{
 \alignauthor Akash Mankar\\
   \affaddr{National Instruments}\\
   \affaddr{Austin, TX USA}\\
   \email{akash.cs@utexas.edu}\\
 \alignauthor Riddhi J Shah\\
   \affaddr{Rackspace}\\
  \affaddr{Austin, TX USA}\\
   \email{ridhi.j.shah@gmail.com}\\
 \alignauthor Matthew Lease\\
   \affaddr{School of Information}\\
   \affaddr{University of Texas at Austin}\\
   \email{ml@utexas.edu}\\
}

\maketitle

\begin{abstract}
Entry-level crowd work is often reported to pay less than minimum wage. While this may be appropriate or even necessary, due to various legal, economic, and pragmatic factors, some Requesters and workers continue to question this status quo.  To promote further discussion on the issue, we asked Requesters and workers if they would support restricting tasks to require minimum wage pay.  As a form of {\em design activism}, we confronted workers with this question more directly by posting a dummy Mechanical Turk task which told them that they could not work on it because it paid less than their {\em local} minimum wage, and we invited their feedback. Strikingly, for those workers expressing an opinion, two-thirds of Indians favored the policy while two-thirds of Americans opposed it. Though a majority of Requesters supported minimum wage pay, only 20\% would enforce it. To further empower Requesters, and to ensure that effort or ignorance are not barriers to change, we provide a simple public API\footnote{\url{https://github.com/akash-mankar/DesignActivism}} to make it easy to find a worker's local minimum wage by his/her IP address. A short version of this article appeared\,at \sc{aaai hcomp} 2017\footnote{\scriptsize \url{http://www.humancomputation.com/2017/}}.

\end{abstract}



\section{Introduction}

Crowdsourcing is driving a new generation of intelligent systems for solving tasks that could not be effectively automated in the past~\cite{Kittur13} and powering new scales of human subjects research~\cite{mason2012conducting}. While a great deal of research has focused on optimizing technological approaches to improving work quality, affordability and speed, fewer studies have wrestled with questions of what constitutes ethical, legal, and sustainable 
practices in online crowdsourcing (cf.~\cite{felstiner2010working,irani,Kittur13,silberman}). 

For Amazon Mechanical Turk (AMT), while some workers report accepting tasks for disposable income or to pass the time, Ross et al.~\cite{Ross:2010:CSD:1753846.1753873} estimated that nearly 27\% of Indian and 14\% of American workers were dependent on AMT income for basic needs. Probing workers' {\em reservation wage} (the smallest amount a worker is willing to accept), Horton et al.~\cite{horton} found an AMT median value of \$1.38/hour.  In the US, while federal minimum wage is \$7.25/hour~\cite{wikipedia}, classification of crowd work as independent contracting excludes it from employment regulation~\cite{felstiner2010working}. Moreover, with greatly varying local minimum wages (e.g., only \$0.28/hour in India~\cite{wikipedia}) and outsourcing practices long-established, are notions of {\em equal pay for equal work} simply antiquated in a 21st century, global and digital economy?  Ethical pay rates for global participants in online studies is also unclear~\cite{mason2012conducting}.

\begin{figure}[ht]
\centering
\ifpdf
\includegraphics[width=0.5\columnwidth]{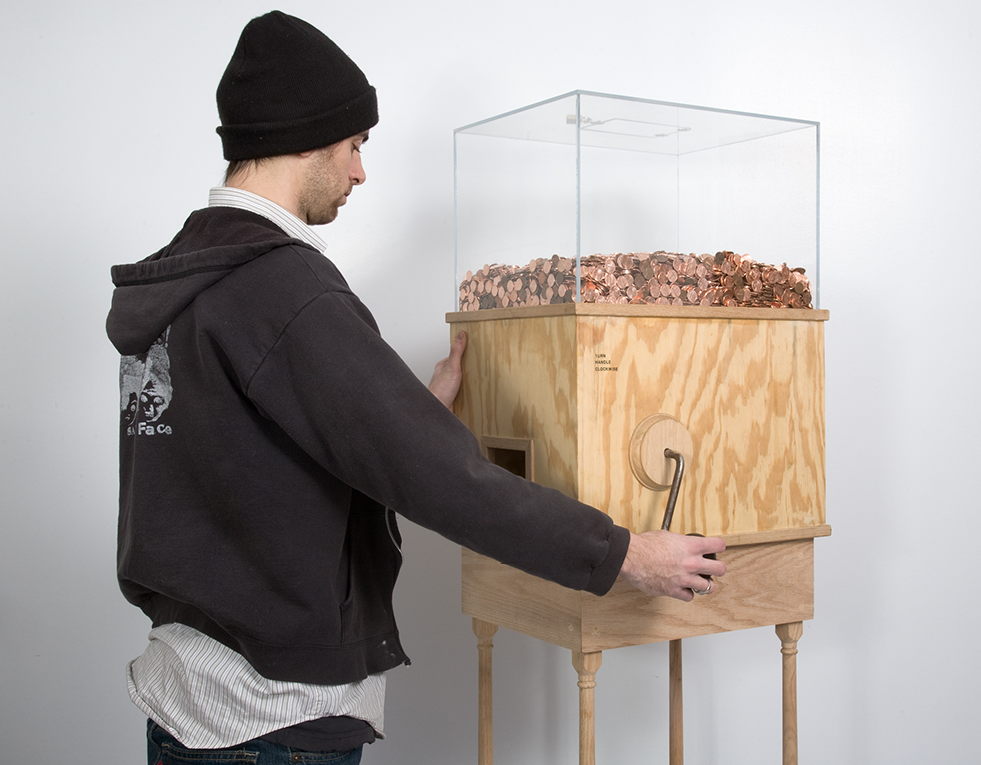}
\fi
\caption{Blake Fall-Conroy's {\em Minimum Wage Machine}.}
\label{fig:machine}
\end{figure}
As {\em design activism}, Turkopticon was deployed to both provide functional enhancements on AMT, helping mitigate power disparities, but also to provoke discussion and disrupt common rhetoric of technocentric optimism surrounding crowd-powered systems~\cite{irani}.  Blake Fall-Conroy's {\em Minimum Wage Machine} (\url{blakefallconroy.com/18.html}) provides a similar visual and experiential opportunity to try offline minimum wage work; turning the mechanical crank will yield one penny every 5 seconds, yielding \$7.25/hour (Figure~\ref{fig:machine}). 

While one can enumerate many potential technological, economic, and legal barriers to any proposed change, we might begin first by simply imagining and debating ideas for what a better model of paid crowd work might look like~\cite{Kittur13}. To seed such community {\em ideation}, let us begin by imagining a reality in which paid crowd work were required to compensate workers according to local minimum wage laws.  Would (any) Requesters be in favor of such a policy?  Would (any) workers support being blocked from available tasks that did not meet this wage restriction in their local region?  Would workers in different regions have different views?

Toward this end, we conduct a simple survey of AMT workers and Requesters (we could not survey researchers without sacrificing our anonymity, but plan to do so). To be even more provocative, we created a dummy Mechanical Turk task in which workers were told after accepting the task that they could not work on it because it paid below the worker's {\em local} minimum wage, at which point we invited their feedback (and offered payment for feedback in lieu of the task). Strikingly, for those workers expressing an opinion, two-thirds of Indians favored the policy while two-thirds of Americans opposed it. Though a majority of Requesters supported minimum wage pay, only 20\% reported willingness to enforce it. 

To empower any Requesters favoring the policy, 
and to further ensure effort or ignorance are not barriers to change, we provide a simple public API to support easily finding each worker's local minimum wage by IP address. Just as Apple iTunes made it easier for people to buy music legally, can technological convenience alone yield higher pay on AMT? 

\section{Related Work}  \label{relwork} 

Several researchers have analyzed different forms of virtual work including crowdsourcing and whether they were entitled to minimum payment standards mandated under the FLSA (Fair Labor Standards Act)~\cite{flsa,felstiner2010working}. 
Worshall~\cite{worshall} claims that it is only fair and ethical to cover the workers by federal minimum-wage legislation. Dolan \cite{dolan} describes how Samasource uses microwork to award fair wages to poor workers in countries with a lower minimum wage. 

Silberman et al.~\cite{silberman} presented some of the first commentary on {\em worker invisibility} and ethics considerations. Martin et al.~\cite{martin} recently surveyed and interviewed workers regarding their expectations and opinions regarding crowd work. Kaufmann et al.~\cite{kaufmann} investigated worker motivations in crowdsourcing, with focus on AMT given the typical payments offered are relatively low. They tested their model by surveying 431 AMT workers. Complementing Ross et al.'s study of demographics~\cite{Ross:2010:CSD:1753846.1753873}, Ipeirotis~\cite{demog} found that 50\% of the AMT workforce was American and 40\% was Indian.

\section{Methodology} \label{meth}

We first issued unpaid surveys for Requesters and workers on TurkerNation (\url{www.turkernation.com}), TurkerReddit (\url{www.reddit.com/r/mturk}), and on AMT itself. These surveys tended to receive only minimal response, especially for workers. For brevity, we exclude from further discussion and analysis the handful of unpaid worker responses received.  However, the unpaid AMT Requester survey did receive many responses, as summarized in Table~\ref{tab:table2}. As a known limitation of our methodology, a survey of Requesters as an AMT task is admittedly more likely to be found by workers than Requesters. More generally, we cannot verify that responses actually come from Requesters, and so workers posing as Requesters could impact our tallies. 

{\small
{\bf Question to workers:} \emph{Would you prefer it if you were allowed to work only on those tasks whose estimated reward was equal to or more than the minimum wage of your country, and be blocked from working on other lower paid tasks? Please give your opinion in at least 3 sentences.}

{\bf Requester Survey}
\begin{enumerate}
\item If you are a requester on a crowd-sourcing platform,would you prefer to reward the worker as per the minimum wage norms of the country the worker belongs to?  (Yes/No/Neutral)
\item If yes, why? Do you think you would get quality results if you did so?
\item If yes, would you block the workers whose minimum wage is greater than what you can offer per person? Why or why not?
\item If no, what are the reasons why you would not prefer this policy? Please state it clearly in 2-3 sentences.
\item Any other comments about the policy?
\end{enumerate}
}
Beyond soliciting such ``arm-chair'' opinions, we also experiment with real-time simulation of the proposed policy by putting workers into a real-life situation in which this policy is enforced. We created attractive tasks on MTurk which ask users to go to a particular link/website to gather certain information, e.g., entering restaurant information by visiting a particular link included in our HIT. In fact, the link actually points to our own website hosted on Google App Engine. 
Java and JSP (Java Server Pages) are used for back-end and front-end design of the website respectively. When the worker visits our website for gathering restaurant information, we lookup the worker's geographic region via IP address using an open source API called GeoIP (\url{freegeoip.net}).

In a real implementation, 
we would imagine that a requester could automatically lookup the appropriate local minimum wage for that worker's region~\cite{wikipedia}. The worker could be restricted from working on the task if it did not pay at least that hourly wage (assuming an estimated average task duration, perhaps by initial guestimate and/or historical task times).

For our experimental data collection, however, we simply told every worker that the task paid too little for them to work on it.  To support believability, we posted different paying versions of the same task for Indian and American workers, with each task restricted to the corresponding region (the minimum wage in India is only \$0.28 per hour). Given that AMT pays local currency in only these regions, we focused on only them for this study; future work might consider additional regions as well. Because we had separate restricted tasks for each region, this also let us easily analyze responses by region.  

Our website displayed an apology, explained the policy to workers, and then invited their feedback, offering one-time payment for that feedback instead. Workers who provided feedback were provided payment in lieu of the actual task.


{\small

\emph{We see you are located in XXX where the local minimum wage is YYY (Source:Wikipedia), and unfortunately the expected completion time for this task would mean you would be paid less than minimum wage. We are not comfortable with that, and cannot afford to pay more, and so are not allowing this task to be performed in your region. We apologize for the inconvenience and welcome your feedback. Do you think this is a good policy that we do not allow our Mechanical Turk workers to work for less than minimum wage, even though that means some tasks we post will be unavailable to workers in certain regions? 
To be fair, if you complete this feedback form and submit it, on the next page, 
you'll be able to find a unique number, which you can still use to complete this HIT and we would accept your task and pay you the amount we promised for this HIT on this form completion. Just copy paste it in the unique code section text box and submit the HIT.}
}

To deter any trivial spam or carelessness, the feedback form was scripted to require completion of all fields. We later report number of visits vs.\ feedback responses received. 

Workers were required to have previously completed 100 HITs in order to access our tasks. We did not require any prior approval rate. Inspired by {\em Turkalytics}~\cite{heymann2011turkalytics}, we also used IP tracking to check if we received visits from the same IP address for different worker IDs, or the same worker ID for different IP addresses. Beyond fraud, this might arise in benign settings like an internet cafe with dynamic IP addressing. 
We detected no such occurrence in practice.  A technological hurdle we could imagine in practice with such a policy is IP-spoofing: if pay is region dependent, one might try to change one's IP to appear to be located in a higher-paying region.

\section{Results} \label{results}

Worker responses are summarized in Table~\ref{tab:table1}. While paid AMT survey responses show workers (regions unknown) as evenly divided on the policy, worker feedback from our AMT dummy task was quite different. 241 of 301 US workers who visited our website left feedback.  Interestingly, we received far fewer Indian visits to our website: 104, with 62 responses. Far more strikingly, the distribution of opinions between regions diametrically-opposed, with two-thirds of Indian workers expressing an opinion in favor of the policy, while two-thirds of such American workers oppose the policy. 

\begin{table}[ht]
\centering
\caption{Worker Survey Results show opposite USA and India responses.}
\begin{tabular}{|l|c|rr|rr|rr|}
\hline
Region & Total & \multicolumn{2}{c|}{Yes} & \multicolumn{2}{c|}{No} & \multicolumn{2}{c|}{Neutral} \\ 
\hline
\multicolumn{8}{|c|}{{\footnotesize\em Paid Survey on Mechanical Turk}} \\
\hline
All & 73 & 32 & 44\% & 33 & 45\% & 8 & 11\%\\  
\hline
\multicolumn{8}{|c|}{{\footnotesize\em Feedback on Mechanical Turk Posted Task}} \\
\hline
USA   & 241 & 55 & {\bf 23\%} & 121 & {\bf 50\%} & 65 & 27\%\\
India & ~~63& 31 & {\bf 49\%} &  15 & {\bf 24\%} & 17 & 27\%\\
All   & 304 & 86 & 28\% & 136 & 45\% & 82 & 27\%\\
\hline
\multicolumn{8}{|c|}{{\footnotesize\em Combined Results}} \\
\hline
All & 377 & 118 & 31\% & 169 & 45\% & 90 & 24\%\\
\hline
\end{tabular}
\label{tab:table1}
\end{table}

Requester feedback is summarized in Table~\ref{tab:table2}. We received a total of 88 responses (82 from AMT, and only 6 from both TurkerNation and TurkerReddit). Across all responses, a surprising majority 58\% of Requesters report favoring minimum wage payment, with 39\% opposed (Question 1). Of those in favor, 71\% expected better quality work would result from such a payment policy, with only 1\% opposed (Question 2). Despite this, only 20\% of Requesters indicated willingness to block workers from performing tasks that would underpay vs.\ local minimum wage laws: 35\% opposed the policy, while a large 45\% of Requesters expressed neutrality on the question.


\begin{table}[ht]
\centering
\caption{Results of Requester Surveys (all unpaid).}
\begin{tabular}{|l|c|rr|rr|rr|}
\hline
\!\!Question\!\! & \!\!Total\!\! & \multicolumn{2}{c|}{Yes} & \multicolumn{2}{c|}{No} & \multicolumn{2}{c|}{Neutral\!\!} \\ 
\hline
\multicolumn{8}{|c|}{{\footnotesize\em TurkerNation \& TurkerReddit}} \\
\hline
\!\!1.\ {\footnotesize Prefer pay min-wage?}\!\!     & \!\!~6\!\! & 2\!\! & & 3\!\! & & 1\!\! & \\  
\!\!2.\ {\footnotesize Expect quality results?}\!\!  & \!\!~2\!\! & 2\!\! & & 0\!\! & & 0\!\! & \\  
\!\!3.\ {\footnotesize Block if pay too low?}\!\!    & \!\!~2\!\! & 0\!\! & & 2\!\! & & 0\!\! & \\  
\hline
\multicolumn{8}{|c|}{{\footnotesize\em Mechanical Turk}} \\
\hline
\!\!1.\ {\footnotesize Prefer pay min-wage?}\!\!    & \!\!82\!\! & 49\!\! & \!\!60\% & 31\!\! & \!\!38\% &  2\!\! &  \!\!2\%\!\!\\
\!\!2.\ {\footnotesize Expect quality results?}\!\! & \!\!49\!\! & 34\!\! & \!\!69\% &  1\!\! &  \!\!2\% & 14\!\! & \!\!29\%\!\!\\
\!\!3.\ {\footnotesize Block if pay too low?}\!\!   & \!\!49\!\! & 10\!\! & \!\!20\% & 16\!\! & \!\!33\% & 23\!\! & \!\!47\%\!\!\\
\hline
\multicolumn{8}{|c|}{{\footnotesize\em Combined}} \\
\hline
\!\!1.\ {\footnotesize Prefer pay min-wage?}\!\!    & \!\!88\!\! & 51\!\! & \!\!58\% & 34\!\! & \!\!39\% &  3\!\! &  \!\!3\%\!\!\\
\!\!2.\ {\footnotesize Expect quality results?}\!\! & \!\!51\!\! & 36\!\! & \!\!71\% &  1\!\! &  \!\!2\% & 14\!\! & \!\!27\%\!\!\\
\!\!3.\ {\footnotesize Block if pay too low?}\!\!   & \!\!51\!\! & 10\!\! & \!\!20\% & 18\!\! & \!\!35\% & 23\!\! & \!\!45\%\!\!\\
\hline
\end{tabular}
\label{tab:table2}
\vspace{-5pt}
\end{table}

\section{Worker Opinions}

In addition to the above tabular data, it was also interesting to note the diversity of open-ended comments we received.  We select a few of these comments to include here as illustrative. The entire dateset is publicly accessible$^\textnormal{1}$. 


\subsection{Supporting Opinions}
\emph{``It is a good policy because it shows you care about the turkers. Also it stops us from wasting our time. This way more people can take part in surveys.''} 

\emph{``\ldots this is a pretty good policy because if you make over a certain amount of dollars on mturk you have to file that information in your taxes. If I am being taxed on wages earned on mturk then shouldn't the minimum wage law apply too?''}

\emph{``This is fantastic because most requesters are fine with having workers work for far less than minimum wage.''}

A recurring theme among positive responses is a sense that being able to get paid minimum wage shows appropriate respect towards workers as a task force. While some workers express concerns that Requesters are often unfair in terms of wage, such a policy might bring greater fairness and equality to all parties.  As shown above, some express a belief that taxation should entitle minimum wage guarantees.


\subsection{Neutral Opinions}
\emph{``I think it is a nice idea but this is something that many people choose to do in their free time and I don't think you need to pay minimum wage.''}

\emph{``I like it because I think it's a very noble idea. The MTurk site truly does not pay very much. However, a lot of us are here because we really need money, and the small amounts add up to something at the end of the month.''}

Some workers in this group suggested that this policy seems ideal but not realistic, while others noted AMT is only contract work and something done only for supplemental income. 


\subsection{Opposing Opinions}
\emph{``If this were not the policy I could still make \$5.00 which I would be fine with and instead I am getting a lot less money than that because I cannot work on the hit. Therefore, I find this policy a bad policy.''}


\emph{``Because I am working from home a few minutes at a time, I don't believe this type of work should come under the fair wage and hour laws. I am making a conscious decision to work on the pay scale provided, although it is low, which I believe is my [right]\ldots''} 

\emph{``This is a bad policy. Minimum wage does not apply to contract workers. Give me my HIT!!!''} 

\emph{``If I'm willing to work for less, I should have the option. I don't think it's up to someone else to decide my fair wage. Besides, my completion time may be quicker, resulting in a higher pay/hour.''}

Workers against the policy expressed firm beliefs that they are capable of making their own decision about whether to attempt a given task. In some cases, workers are happy getting paid less than minimum wage because the amount is still significant to them, and such a policy could prevent this. 

\section{Requester Opinions}

\subsection{Supporting Opinions}
\emph{``The minimum wage seems like the most fair way to reward people. I would not exclude people who usually make a higher minimum as long as they know what I was offering.''}

As noted earlier, a majority of Requesters supported the minimum wage policy. Their comments suggest a firm belief that it is a fair way to reward people working on AMT tasks. Some of these opinions were argued based on standards and ethics. 

\subsection{Neutral opinions}
\emph{``It depends on the task and the difficulty involved I wouldn't block someone, unless they broke the rules I set up''} 

Few of the opinions were neutral, but interesting. They suggested that these Requesters were fine with paying the minimum wage assuming quality improvement justified it. However, they did not want to block anyone.

\subsection{Opposing Opinions}
\emph{``It is unfair to give people minimum wage [by] country when this is a crowd sourced platform where people come from different parts of the world. They deserve to be treated equally.''} 

Such requesters suggested that internet as a whole is one entity, and that people are ``netizens'' of the Internet and not citizens of any particular country. In the spirit of {\em equal pay for equal work}, these comments suggested all netizens should be treated and paid equally, irrespective of physical location.

\section{Further Analysis of Feedback}
In reviewing opinions we received, we further organized them into emergent categories to identify recurring themes and their relative distributions. Results appear in Table~\ref{tab:table4}.
We exclude neutral opinions, which did not seem to have any distinctive categories. The most common reason for a neutral opinion for both Requesters and workers was principally that their opinion depends on the type of task at question. 

\begin{table}[ht]
\small
\centering
\caption{Sub-categorization of worker and Requester opinions}
\begin{tabular}{|l|l|c|}
\hline
\hline
                                \multicolumn{3}{|c|}{\textbf{Workers}}                                                                              \\ \hline
                                & \multicolumn{1}{c|}{\textit{\textbf{In Favor of the Policy}}}                                                      & \textit{\textbf{118}} \\ \hline
1                               & Ethically Fair                                                                                                     & 42                    \\ \hline
2                               & Legally Compliant                                                                                                  & 8                     \\ \hline
3                               & Help them find higher paying work more quickly                                                                     & 8                     \\ \hline
4                               & Encourages them to do better work                                                                                  & 4                     \\ \hline
5                               & Will help meet their basic end needs                                                                               & 9                     \\ \hline
6                               & Miscellaneous                                                                                                      & 47                    \\ \hline
\multicolumn{1}{|c|}{\textit{}} & \multicolumn{1}{c|}{\textit{\textbf{Against the Policy}}}                                                          & \textit{\textbf{169}} \\ \hline
1                               & \begin{tabular}[c]{@{}l@{}}Ethically Unfair  (performance/merit not location \\ and its minimum wage)\end{tabular} & 11                    \\ \hline
2                               & Opportunities reduce                                                                                               & 18                    \\ \hline
3                               & It is not a MTurk/crowdsourcing suitable policy                                                                    & 19                    \\ \hline
4                               & Not a source of basic income                                                                                       & 14                    \\ \hline
5                               & Workers capable of making a conscious decision                                                                     & 71                    \\ \hline
6                               & Not here for money only/Like working on tasks                                                                      & 9                     \\ \hline
7                               & Miscellaneous                                                                                                      & 27                    \\ 
\hline 
\hline
\multicolumn{3}{c}{}\\
\hline
\hline

                                \multicolumn{3}{|c|}{\textbf{Requesters}}\\ \hline
                                & \multicolumn{1}{c|}{\textit{\textbf{In Favor of the Policy}}}                                                      & \textit{\textbf{51}}  \\ \hline
\multicolumn{1}{|c|}{1}         & Incentive for quality work                                                                                          & 36                    \\ \hline
\multicolumn{1}{|c|}{2}         & Ethically fair                                                                                                     & 8                     \\ \hline
\multicolumn{1}{|c|}{3}         & Attracts more workers                                                                                              & 6                     \\ \hline
\multicolumn{1}{|c|}{4}         & Miscellaneous                                                                                                      & 1                     \\ \hline
                                & \multicolumn{1}{c|}{\textit{\textbf{Against the Policy}}}                                                          & \textit{\textbf{34}}  \\ \hline
\multicolumn{1}{|c|}{1}         & Not a MTurk/crowdsourcing suitable policy                                                                          & 5                     \\ \hline
\multicolumn{1}{|c|}{2}         & Ethically unfair                                                                                                   & 10                    \\ \hline
\multicolumn{1}{|c|}{3}         & Work is low skill/low effort, not worth minimum wage                                                               & 8                     \\ \hline
\multicolumn{1}{|c|}{4}         & Cannot afford or lack of funding                                                                                   & 2                     \\ \hline
\multicolumn{1}{|c|}{5}         & Do not want to spend (more productive / business model)                                                            & 2                     \\ \hline
\multicolumn{1}{|c|}{6}         & Miscellaneous                                                                                                      & 7                     \\ \hline
\hline
\end{tabular}
\label{tab:table4}
\end{table}

\section{Empowering Requesters} \label{emp}

As noted earlier, a majority of Requesters supported our posited minimum wage policy. In order to empower this section of the Requester community, we created a simple Python API$^\textnormal{1}$
which can be used to help implement this policy in practice. 
Given the worker's IP address, the Requester can first lookup the worker's region via the aforementioned open-source GeoIP library. Next, the Requester uses our API to automatically look up the local minimum wage for that region, using rates from Wikipedia~\cite{wikipedia}.


As an example, imagine a Requester wants to publish a task in India with 10 HITs and allocates 6 minutes for the completion of each HIT (60 minutes for all the HITs). Before specifying the reward, the requester runs the python program and provides the location (India) to it as a command line argument. This returns the minimum wage/hour in US\$ for the country India (\$0.28/hour). If the requester wants to enforce the minimum wage policy, he/she has to make sure the estimated net reward for the task is greater than \$0.28/hour, i.e., at least \$0.028 per HIT. For a general task available globally, the Requester could dynamically detect the worker's region and either block the worker from performing the task, or provide bonus payment up to the local minimum wage rate.



\section{Conclusion and Future Work} \label{conc}

We conducted this study without an intent to insist that our ``straw-hat'' policy would be an ideal and universal solution to fairness in crowdsourcing. Instead, we built a system whose results are useful to make worker---requester relationship more transparent and to provoke the thoughts of globally situated workers and Requesters for an ethical debate regarding minimum wage. The responses that we received reflect interesting patterns of thought for engaging dialog and informing researchers and others seeking to imagine alternative future designs for ethical and sustainable crowdsourcing.


We would like to advertise our API to, and solicit feedback from, the research community, to further our own understanding and to promote ongoing dialog.

\section*{Acknowledgements} 

We thank our talented crowd workers, without which our study would not be possible, as well as fellow researchers who provided valuable feedback in preparing this paper. This study was supported in part by National Science Foundation grant No. 1253413 and IMLS grant RE-04-13-0042-13. Any opinions, findings, and conclusions or recommendations expressed by the authors are entirely their own and do not represent those of the sponsoring agencies.


\bibliographystyle{acm-sigchi}
\bibliography{sample}





%
%
%
%
%

\end{document}